\date{}
\documentclass[twocolumn,showpacs,superscriptaddress]{revtex4}
\usepackage{graphicx,psfrag,amsmath,amssymb,amsfonts,bbm,latexsym,color,dcolumn,
epsf,graphpap}

\definecolor{red}{rgb}{1,0,0}
\definecolor{blue}{rgb}{0,0,1}
\definecolor{skyblue}{rgb}{0,0,.5}
\definecolor{green}{rgb}{0,1,0}
\definecolor{orange}{cmyk}{0,.4,1,0}

\begin{document}
\title{Decoherence, tunneling and noise-activation in a double-potential well 
at high and zero temperature}

\author{Nuno D. Antunes \footnote{n.d.antunes@sussex.ac.uk}}
\affiliation{Centre for Theoretical Physics, University of Sussex,
Falmer, Brighton BN1 9QJ - UK}
\author{Fernando C. Lombardo \footnote{lombardo@df.uba.ar}}
\affiliation{Departamento de F\'\i sica {\it Juan Jos\'e Giambiagi}, FCEyN UBA,
Facultad de Ciencias Exactas y Naturales, Ciudad Universitaria,
Pabell\' on I, 1428 Buenos Aires, Argentina}
\author{Diana Monteoliva \footnote{monteoli@df.uba.ar}}
\affiliation{Departamento de F\'\i sica, Facultad de Ciencias Exactas de la Universidad
Nacional de La Plata, La Plata, Argentina}
\author{Paula I. Villar \footnote{paula@df.uba.ar}}
\affiliation{Departamento de F\'\i sica {\it Juan Jos\'e Giambiagi}, FCEyN UBA,
Facultad de Ciencias Exactas y Naturales, Ciudad Universitaria,
Pabell\' on I, 1428 Buenos Aires, Argentina}

\date{today}

\begin{abstract}

We study the effects of the environment on tunneling in an open system 
described by a static  double-well potential. We describe the evolution
of a quantum state  localized in one of the minima of
the potential at $t=0$, both in the limits of high and zero environment
temperature. We show that the evolution of the system can be summarized in 
terms of three main physical phenomena, namely decoherence, quantum tunneling
and noise-induced activation, and we obtain analytical estimates for the
corresponding time-scales. These analytical predictions are confirmed by 
large-scale numerical simulations, providing a detailed picture of the 
main stages of the evolution and of the relevant dynamical processes.

\end{abstract}


\maketitle

\newcommand{\beq}{\begin{equation}}
\newcommand{\eeq}{\end{equation}}
\newcommand{\dalam}{\nabla^2-\partial_t^2}

\section{Introduction}
\label{intro}

The emergence of classical behavior in quantum systems is a topic of
great interest from both conceptual and experimental points of view \cite{exp}.
It is well established by now that the interaction between a quantum
system and an external environment can lead to its classicalization;
decoherence and the occurrence of classical correlations being the main
features of this process (for a recent overview see \cite{Paz_Zurek}).

One of the most intriguing prospects in quantum physics 
is the possibility of observing quantum tunneling on macroscopic scales \cite{caleg,gral}.
Macroscopic systems are generally open systems, interacting with an external 
environment, and in this context quantum tunneling 
is qualitative different from its experimentally verified microscopic analogue \cite{hanggi}.

The analysis of open systems has led to interesting results, detailing
the dynamics of a quantum system coupled to a thermal
bath with arbitrary temperature. A closed quantum system described by a
state localized around a meta-stable minimum, should tunnel through the
potential barrier with a well defined time-scale. This tunneling time can be
estimated using standard techniques such as the instaton method
\cite{Coleman}. For an open system, on the other hand, it is well known
that the environment induces decoherence on the quantum particle,
its behavior becoming classical as soon as interference terms are destroyed
by the external noise \cite{phz}. This transition from a quantum to a classical 
behavior is forced by the interaction with a  robust environment and takes place 
at a given time-scale, the decoherence time \cite{zurek}. This quantity depends on 
the properties of the system, its environment and their mutual coupling. 
If the decoherence time is 
significantly smaller than the tunneling time, one would expect that
after classicalization the state should become confined to the meta-stable 
vacuum, with tunneling being suppressed. The particle could still cross
the barrier but only if excited by the bath, its energy increasing via 
thermal activation, for example. This process is distinct from
quantum tunneling, is classical in its nature and should be efficient
mostly at high environmental temperatures.

An interesting question arises: what is the effect on tunneling if
the particle is coupled to a reservoir at zero temperature? 
Though in this case there should be in principle no thermal activation,
we know there is decoherence induced by a quantum environment at zero 
temperature \cite{Paz_Zurek,davila,pau,butt}. This would lead to 
classicalization and one could conclude that even at $T=0$, quantum tunneling should
be inhibited by the interaction with the external environment \cite{calzetta1}.

The study of the effects of an external environment on tunneling
was initiated by Caldeira and Legget \cite{caleg} who showed that 
dissipation inhibits tunneling.
Dissipation in the primary system is not the only result of the interaction 
with the external environment. Not only does the bath lead to the renormalization 
of the coupling constants and of the frequency but it dissipation induces 
fluctuation (noise) on the system as well \cite{qbm}. Since this ground-breaking 
study \cite{caleg,leggett2}, many other works
have looked at the various aspects of the same phenomenon arriving 
more often than not at similar conclusions \cite{gral2}. What's even more appealing 
is the fact that nearly all studies rely 
on analytical techniques, either functional based approaches or
generalizations of instanton-type calculations. These approaches are 
based on equilibrium concepts and may miss important dynamical aspects
of the evolution. In particular, it is hard to treat 
both tunneling and activation-like effects simultaneously, and to differentiate 
their individual contribution to the outcomes.

Since the early works on open quantum systems, computational power
has increased hugely and it is now feasible to test and extend
many analytical results using large-scale numerical simulations.
In the context of simulations of open-systems with tunneling effects,
the main focus has been on
driven systems, where tunneling between regular and non-regular islands is
of interest \cite{tunel,diana_jpp}. In these models, the interplay
of classical chaos and dissipation bears interesting 
effects at the boundary between classical and quantum mechanics, e.g., the
elimination of classical chaos by quantum interference, or its
restoration by dissipation. 

In this article we will concentrate on a simple tunneling  system 
described by a static Hamiltonian. Specifically, we will look in detail at 
the evolution of a particle in
a quantum state localized at one of the minima of a double well potential, 
when coupled to an external environment at both zero and high temperature.
We will present an analytical description of the effects of dissipation and
diffusion, and estimate the time-scales associated with the 
distinct physical processes governing the dynamics
of the system: decoherence, quantum tunneling and activation.
Using large-scale numerical simulations we will then be able to obtain
a full description of the dynamics of the model, and test the analytical
estimates. Moreover, since we will have access to the state of the system
at any time, we will be able to distinguish between the effects of the
several processes mentioned above.

As we will discuss below, we confirm that in the high-temperature regime,
the evolution can be indeed 
well understood in terms of simple decoherence, tunneling and activation
time-scales. This enlightens the problem conceptually and offers a
great degree of control over the behavior of the system. In particular, we
will see how the environment can be manipulated in order to delay 
or accelerate decoherence, and how the strength of the bath's coupling allows
activation to be retarded. The estimates provided can be used in wider 
situations and hopefully be generalized to realistic systems with 
tunneling on macroscopic scales. Finally, we will show than in
the particular case of a zero temperature environment, not only
tunneling is inhibited, but
contrarily to what may be naively expected, noise-activation is
also observed. Consequently, 
at large times after decoherence, the particle has always a non-zero
probability of crossing the potential barrier. We will discuss how this
 result can be 
understood by means of a the classical finite-temperature analogue.

The paper is organized as follows. In Section II we present our
model and derive estimates of the relevant scales involved. In Section III we
analyze the high temperature limit, showing how
 decoherence inhibits tunneling 
and describing thermal activation in the classical regime. This is done using both analytical 
and numerical results. Section IV contains a similar analysis detailing the 
case of zero environmental temperature. Finally, in Section V,  we include our final
remarks and in the Appendix we expand on more technical details of calculations 
relevant for the body of the paper.

\section{The model}
\label{model}

We will start by considering a quantum anharmonic oscillator
with a potential given by $V(x)=-\frac{1}{4}\Omega^2 x^2 + \lambda x^4$. 
This is a double
well potential with two absolute minima at 
$x_0=\pm \Omega/\sqrt{8\lambda}$
 separated by a potential barrier with height 
$V_0=\Omega^4/(64\lambda)$\textcolor{blue}.
We will assume that the system is open, meaning that
it is coupled to an environment composed of an infinite set of harmonic
oscillators \cite{diana_mazzi}.
The complete classical action for the system and environment is given by:

\begin{eqnarray}S[x,q_n] &=& S_{\rm sys}[x] + S_{\rm env}[q_n] +
 S_{\rm int}[x,q_n]\nonumber \\
&=& \int_0^t ds \left[ \frac{1}{2}{\dot x}^2 + \frac{1}{4}\Omega^2 x^2 -
\lambda x^4 \right. \nonumber \\
&+& \left. \sum_n {1\over{2}} m_n ({\dot q}_n^2 - \omega_n^2 q_n^2)\right] - 
\sum_n C_n x q_n,\label{action} \end{eqnarray}  
where $q_n$, $m_n$ and $\omega_n$  are respectively the coordinates, 
masses and frequencies of the environmental oscillators. The mass
of the anharmonic oscillator is set to one.
The main system is coupled linearly to each oscillator in the
bath with strength $C_n$. 
This action describes one of the most simple quantum Brownian
 motion (QBM) 
models, which has been widely used in the study of quantum to
 classical transition phenomena \cite{phz,qbm}. 

The dynamics of the non-linear oscillator can be obtained by
tracing over the degrees of freedom of the environment and
obtaining a master equation for the reduced density matrix of the system,
$\rho_{\rm r}(t)$. We will assume that the initial states of the system and
environment are uncorrelated, with the latter being in thermal 
equilibrium at temperature $T$ (possibly zero) for $t=0$ (i.e when the 
interaction between system and environment is switched on). At the 
initial time, the state is a product of a given state of system (entirely on 
the left well) and a thermal state for the environment. Only when the 
interaction is turned on the system is allowed to evolve. The initial condition
is not an equilibrium state of the complete action. Under these assumptions, and using 
that the system-environment coupling is small, the reduced
density matrix satisfies the following time-convolutionless master equation \cite{convo}:

\begin{eqnarray}
\dot \rho_{\rm r}(t)&=& -i [H_{\rm sys}, \rho_{\rm r}(t)] \nonumber \\
&&-
\int_0^t d\tau \left\{ \nu(\tau) \,[x(t),[x(-\tau),\rho_{\rm r}(t)]]
\right. \nonumber \\
&& \left. - \imath \;
 \eta(\tau) \, [x(t),\{x(-\tau),\rho_{\rm r}(t) \}] \right\}.
\label{master}
\end{eqnarray} 

This equation is perturbative, and therefore we will work with a reduced density 
matrix which is obtained in second order of the system-environment coupling. This fact 
will be taken into account in all the simulations we will present. We will work in
the under-damped case, which ensures the validity of the 
perturbative solutions up to the times we are interested in \cite{qbm,diana_jpp}.
 Therefore, we will keep in mind that we are considering 
an ohmic environment and consider  
situations in which $\gamma_0 \ll \hbar$. This is a weakly-interacting 
setting and from it one can define the temporal domain of validity for
 perturbative solutions. All the results obtained below are for periods
of the evolution well within the regime for which this approximation
is valid \cite{takagi}. $H_{\rm sys}$ is the Hamiltonian for the closed system and 
$x(t) = e^{i H_{\rm sys} t} \, x \, 
e^{- i H_{\rm sys} t} $ the position
operator in the Heisenberg picture. Here and in the following, we
will work in units of  $\hbar=1$. 
$\eta$ and $\nu$ are the dissipation and noise kernels respectively, 
defined as

\begin{eqnarray}\eta (t)& =& \int_0^\infty d\omega I(\omega ) \sin \omega t 
\label{kernel1} \\
\nu (t) &=& \int_0^\infty d\omega I(\omega ) \coth {\beta
\omega\over{2}} \cos \omega t, \label{kernel2}\end{eqnarray} 
where 
$I(\omega )= \sum_n C_n^2 \frac{\delta(\omega - \omega_n)}{2m_n \omega_n}$
 is the spectral density of the environment and
$\beta=1/T$ its inverse temperature (with Boltzmann constant
set to unity, $k_B=1$). It is worth noting that Eq.~(\ref{master}) 
is valid at any temperature, and is local in time, despite the
fact that no Markovian approximation was explicitly taken. In the
next few sections, we will show how the general master equation
simplifies in different regimes, making it more tractable for both 
analytical and numerical techniques. 

As discussed in Section \ref{intro}, we are interested in studying 
tunneling-like phenomena. With this in mind, we will
look at the evolution of a state for which the particle is initially
localized in one of the sides of the double potential well.
In particular, we take as initial condition for the main system
a Gaussian wave function centered around 
the left-hand minimum of the potential, $x_0=-\Omega/\sqrt{8\lambda}$:
\begin{equation}
\Psi_0(x)=\frac{1}{(2\pi\sigma_x^2)^{1/4}} 
            \exp{\left[ -\frac{(x-x_0)^2}{4 \sigma_x^2}\right]}.
\label{psi0}
\end{equation}
The width of the Gaussian is set to $\sigma_x=1/\sqrt{2\Omega}$,
corresponding to the vacuum state for a harmonic oscillator with
frequency $\Omega$. At this point we should take into account that 
once the main system is coupled to the environment, the oscillator 
changes its frequency to a shifted one due to the coupling. We will 
set parameters in order that this frequency shift can be neglected at
all times (we will come back to this point below, when we discuss numerical 
results at zero temperature).  
The frequency $\Omega$ is the natural frequency obtained
by expanding $V(x)$ in the vicinity of its minimum $x_0$ with $\Psi_0(x)$, 
thus, it describes a particle which is ``locally'' in the ``vacuum''.
For the closed system, we expect the state to tunnel through the 
potential barrier: after a tunneling time $\tau$, the wave function
should be approximately given by a Gaussian with similar width centered
on the right-hand minimum of the potential. The tunneling time can
be estimated using standard techniques. The initial Gaussian is well
approximated by a linear combination of the first two energy eigenstates
of the full potential $V(x)$. Denoting the energies of the 
symmetric/anti-symmetric eigenstates by $E_0$ and $E_1$ respectively,
we expect the tunneling time to be given by $\tau \simeq 1/(E_1-E_0)$.
However, as the initial condition Eq.(\ref{psi0}) is not an exact sum of the two
eigenstates, there will be a correction in the tunneling time.
Numerically, as discussed below, we found that in general $\tau=3./(E_1-E_0)$.
 The energy difference and corresponding tunneling time can be obtained
 by a straightforward instanton
calculation \cite{Coleman}, the final result being:
\begin{equation}
 \tau=\frac{3.}{E_1-E_0}=
\frac{3}{8}\sqrt{\frac{\pi}{2}\frac{\Omega}{V_0}}
\frac{1}{\Omega} \exp{\left[\frac{16}{3}\frac{V_0}{\Omega}\right]}.
\label{initial}
\end{equation}
 The expression inside the exponential is 
 the classical action for the instanton, $S_0=(16/3)\times V_0/\Omega$.

\section{Tunneling inhibition at high-T}

At high temperature the reduced master equation can be expressed
in a much simplified way by means of the (also reduced) 
Wigner distribution function on phase space $W=W(x,p;t)$\cite{Paz_Zurek,phz}:

\begin{eqnarray}\dot W &=& \{H_{\rm sys},W\}_{\rm PB}
- {\lambda\over{4}} x 
\partial^3_{ppp}W\nonumber \\
&+&2\gamma (t) \partial_p(pW) + D(t) \partial^2_{pp}W 
- f(t) \partial^2_{px}W,
\label{wigner_eq}
\end{eqnarray}
where

\begin{eqnarray}
\gamma(t) &=& -{1\over 2\Omega}\int_0^t d\tau\sin(\Omega \tau) 
\eta(\tau) \label{gamma},
\\
D(t) &=& \int_0^t d\tau\cos(\Omega \tau) \nu(\tau), \label{D} \\
f(t) &=& -{1\over \Omega}\int_0^t d\tau\sin(\Omega \tau)
 \eta(\tau)\label{f}.
\end{eqnarray}
$\gamma (t)$ is the dissipation coefficient, $D(t)$ and 
$f(t)$ are the diffusion coefficients, all of them given in
terms of the dissipation and noise kernels defined in Eqs.~(\ref{kernel1}-\ref{kernel2}).
The first term on the right-hand side of Eq.(\ref{wigner_eq}) is the
Poisson bracket, corresponding to the usual classical evolution. The second 
term includes the quantum corrections to the dynamics. The last
three terms describe dissipation and diffusion effects due to the coupling to the 
environment. 
In order to simplify the problem, we consider a high-temperature
Ohmic environment, {\it i.e.} we take  $I(\omega )= \frac{2}{\pi}
\gamma_0 \omega \frac{\Lambda^2}{\Lambda^2+\omega^2}$,
where $\Lambda$ is a high frequency cutoff which is larger
than any frequency involved in the system.
In this approximation the coefficients in
Eq.(\ref{wigner_eq}) become constant in time: $\gamma = \gamma_0$, $f\sim 1/T$,
and $D = 2 \gamma_0 T$. The anomalous diffusion coefficient $f$ is much 
smaller than the other ones and therefore we neglect it in Eq.(\ref{wigner_eq}).
 It is important to note that 
the high-temperature approximation is well defined only after a time-scale 
of the order of $1/T\sim\gamma_0/D$. For all cases we will be
studying the relevant period of the evolution lies at much later
times, well in the regime where the approximation holds on.

As discussed in Section~\ref{intro}, the thermal bath will have two 
distinct effects on the evolution of the initial wave packet.
In a regime where the weak coupling to the environment is strong
enough,
the diffusion will make the initial quantum packet decohere, quantum
interference terms will be suppressed and the system will behave classically.
After the decoherence time $t_{\rm D}$, quantum behavior will be
inhibited and tunneling should not be possible any longer. Because the initial 
energy of the particle is less than the barrier height $V_0$, we would
expect it to remain localized on the initial side of the
barrier after $t_{\rm D}$. 
On the other hand, since the particle is in contact with a
high-temperature environment it will ``warm up'' and in time its energy
will increase. At some time $t_{\rm th}$ there will be a significant
probability for the particle to cross through the top of the
barrier, via thermal activation. For very long times, the system
should reach a state of thermal equilibrium, with the particle being
equally likely to be found on either side of the barrier.
     
We will now estimate these two time-scales. In particular, we will be 
interested in understanding how $t_{\rm D}$ and $t_{\rm th}$
interplay with each-other, making the crossing of the barrier more
or less likely at different stages of the evolution.
     
The decoherence time in the high-T limit is usually assumed to be inversely
proportional to the diffusion term $D$ and to the square of the
spatial extension of the wave packet $L$. For our choice of initial
conditions we assume that for early times $L$ can can be set to
the width of the original Gaussian wave function, that is 
$L=2\sigma_x=2/\sqrt{2\Omega}$.
Using $D=2\gamma_0T$ we obtain
(in units of $\hbar=1$) \cite{phz}:
\begin{equation}
t_{\rm D}=\frac{\Omega}{4\gamma_0 T}.
\end{equation}
Though the result is not exact, with $t_{\rm D}$ being slightly overestimated due to the
choice of $L$, its accuracy is enough for our purposes. 

 The thermal activation rate for a classical system can be obtained 
by working with the classical analogue of Eq.~(\ref{wigner_eq}), the
Fokker-Planck equation:
\begin{eqnarray}
\dot{W}=\{H_{\rm sys},W\}_{\rm PB}
+2 \gamma_0 \partial_p(pW) + D \partial^2_{pp}W\,.
\label{fp}
\end{eqnarray}
Note that after decoherence takes place and quantum terms become
irrelevant for the evolution, Eq.~(\ref{wigner_eq}) reduces to 
Eq.~(\ref{fp}). The classical evolution for the average of any
physical observable $A(x,p)$ in this regime is then
given by:
\begin{equation}
\partial_t \langle A \rangle = -\langle\{H_{\rm sys},A\}_{\rm PB} \rangle +
 D \langle \partial^2_p A \rangle - 2 \gamma_0 \langle p
\partial_p A\rangle.
\end{equation}
If we take $A(x,p)$ to be the Hamiltonian of the main system, we
obtain $\partial_t \langle H \rangle = 2 \gamma_0 (T - \langle p^2 \rangle)$.
This expression can be further simplified by assuming $T$ to be much
higher than the relevant energy scales in the problem, $V_0$ and $\langle
p^2 \rangle$, during the
early stages of the evolution. As a result, the time dependence of the
energy of the system is given by:
\begin{equation}
\partial_t \langle H \rangle = 
2 \gamma_0 T \,\,\,\rightarrow\,\,\, E=E_0+2 \gamma_0 T t ~,
\end{equation}
where $E_0$ is the initial energy of the system.
We can then estimate the thermal activation time $t_{\rm th}$
 to be of the same order of the time it takes the system to reach,
 on average, the energy of the height of the barrier:
\begin{equation}
t_{\rm th}=\frac{V_0-E_0}{2 \gamma_0 T}.
\label{t_th}
\end{equation}
When the energy of the initial state is considerably smaller than the potential
height this reduces to $t_{\rm th}=V_0/(2 \gamma_0 T)$. 
These estimates clearly show that there is a large region of parameter
space where it is possible to have decoherence taking place  before the
tunneling time, and delay considerably thermal activation. In these cases
the particle should remain confined in the original side of the barrier
as long as $t<t_{\rm th}$. Ideally we would like to have 
$t_{\rm D}$ and $t_{\rm th}$  separated as much as possible
from the tunneling time scale $\tau$, that is
$t_{\rm D} \ll \tau \ll t_{\rm th}$. For practical purposes we
write:
\begin{equation}
a \, t_{\rm D}  =  \tau  =  b \,t_{\rm th}\,.
\label{tDlltaulltth}
\end{equation}
From the first and last terms we find a restriction on the parameters
of the potential:
\begin{equation}
\frac{V_0}{\Omega} = \frac{1}{2} \left( \frac{a}{b} + 1 \right )\,.
\label{V0Omega}
\end{equation}
Together with a choice of tunneling time, Eq.~(\ref{V0Omega}) fixes
the potential of the main system. The parameters of the environment can
then be set using the first part of Eq.~(\ref{tDlltaulltth}):
\begin{equation}
\gamma_0 T = \frac{a \Omega}{4 \tau}\,.
\label{gamma0T}
\end{equation}
 A choice of $a>>1$ and $b<<1$ would lead to the desired result, keeping
the particle localized one side of the potential well for an arbitrary long time.

\subsection{Numerical Simulation}
\label{numerical}

Our goal here is to use a numerical simulation to test and illustrate the
suppression mechanism discussed above. In terms of  the notation of
 Eq.~(\ref{tDlltaulltth}) 
we should favour a system with large $a$ and small $b$.	Though
such values are perfectly admissible physically, they correspond to
a situation which is hard to tackle numerically. From Eq.~(\ref{V0Omega})
we see that a large ratio of $a$ to $b$ implies a high value for $V_0/\Omega$.
This quantity, $n=V_0/\Omega$, is none other than the semi-classical estimate
for the number of states trapped in the potential well. As we will discuss
below, our numerical method is based on evolving an equation for the eigenstates,
which will be in large number. Since the tunneling time depends exponentially
on $n$, we will also be faced with very large integration times. As a consequence
we will have to chose ``conservative'' values for $a$ and $b$. Nevertheless,
the results will still describe in a conclusive way the phenomena described
in the previous section. 

\subsubsection{Numerical method}

The master equation~(\ref{wigner_eq}) can only be solved by step-by-step
methods up to relatively short times.
As a way out of this problem we resorted to numerically integrating
equation (\ref{master}) on the basis $|\mu\rangle$ of eigenstates of
the Hamiltonian of the isolated
system, $H_{\rm sys} = \frac{p^2}{2} + V(x)$. On our high--$T$ limit this reduces
to:
\begin{equation}
\dot \rho_{\mu\nu} = - \sum_{\alpha\beta} M_{\mu\nu\alpha\beta} \;
 \rho_{\alpha\beta}\,,
\label{rhomunu}
\end{equation}
where $M$ is {\it time--independent}.  The full expression for
$M$ can be found in Appendix.  We have for notation simplicity
dropped the sub-index $\rm r$ on $\rho_{\rm r}$.

As equation (\ref{rhomunu}) has {\it constant} coefficients, it can be
integrated up to {\it any} time once the coefficients $M$ are numerically
calculated.  That means, we can write the
exact solution of the master equation (\ref{rhomunu}) in terms of $M$ as:
\begin{equation}
\rho_{\mu\nu} = \sum_{\alpha\beta} \left({\rm e}^{-M t}\right)_{\mu\nu\alpha\beta} 
\; \rho_{\alpha\beta}(0).
\end{equation}
All the difficulty is now shifted to the calculation of the eigenstates $|\mu\rangle$ of
the system $H_{\rm sys}$, the construction of $M$ and the calculation
of its exponential.  Here problems may arise since $M$ has dimension
$N^4$ (where $N$ is the dimension of the space representing the
real Hilbert space of the problem). Thus $N$ should not be too large.
On the other hand, the number $N$ of states should be large enough to faithfully
represent the system's Hilbert space:  as decoherence couples these states,
the expected quasi-equilibrium state resulting from the master equation
should be diagonal in the eigenstate basis. The master equation
tends to mix these states in such a way that the entropy 
grows to a level where all states become occupied with equal probability.
This provides a good criterion for the validity of the numerical
simulation - in practice we will trust the numerical solution of the
master equation only up to times when the entropy $S$ is below saturation,
{\it i.e.} $S<S_{\rm sat}=\ln N$.

\subsubsection{Decoherence inhibits tunneling}

We have solved equation (\ref{rhomunu}) for the system parameters
$V_0 = 100$ and $\Omega = 5$, which leads to $n = 20$. For this set
of parameters the estimated tunneling time is $\tau = 4.63155403 \, 10^{10}$.
We have chosen $a=24.5$ and $b = 0.6282$ so that $\gamma_0 T = 3.9 \, 10^{-11}$,
which is a very small value. This is to be desired so that the system 
 heats very slowly, delaying thermal activation until after the tunneling
time. Finally, we obtain the relation between the three time scales 
 $t_D \sim 0.0408 \;\tau$ and $t_{\rm th} \sim  1.6326 \;\tau$.

As the initial state is well expanded by 10 eigenstates
of $H_{\rm sys}$ and $n=20$, we have chosen a 
Hilbert space with $N=40$ which is as large a value of $N$ as we
can afford numerically. The environment high frequency cutoff is set
 to $\Lambda = 10 \times \Delta_{40,0} = 10 
\times 102.237307 \gg \Delta_{\alpha\beta}$ for all $\alpha, \; \beta$.
$ \Delta_{\alpha\beta}$ is the frequency separation for eigenstates
$\alpha$ and $\beta$.

\begin{figure}
\epsfxsize=7.6cm
\epsfbox{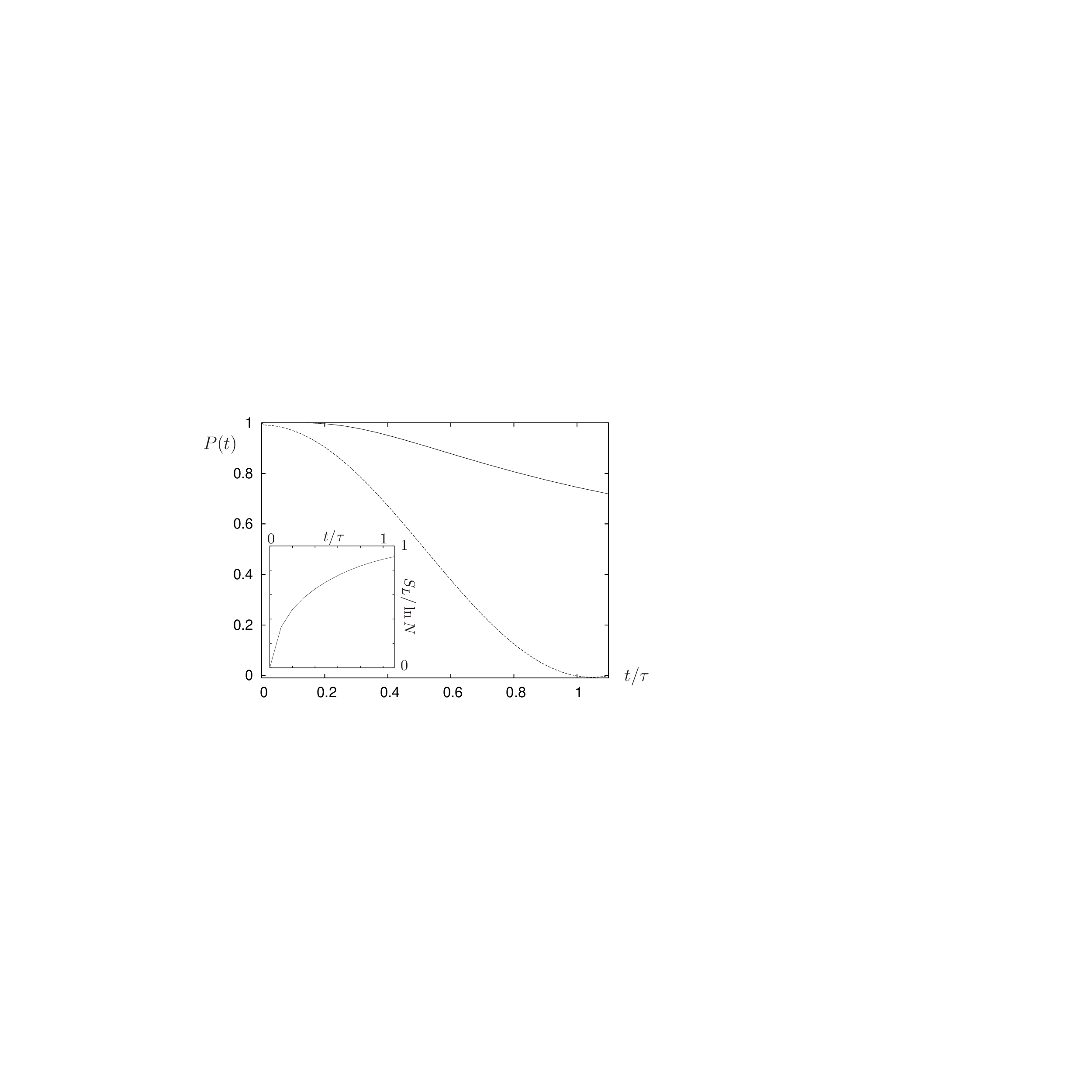}
\caption{Time behavior of the probability to stay on the left of the
barrier for the open (full line) and isolated system (dotted line).
Time is measured in units of the estimated tunneling time $\tau$.
The inset shows, for the same times-pan, the evolution of the linear entropy
of the open system, $S_L/\ln N$ where $N=40$ is the size of the finite space
representing the Hilbert space of the problem.  For $t\sim\tau$ the entropy
is near saturation $S_L/\ln N = 1$ (see text).
\label{figure1}}
\end{figure}

For the numerical solution of the isolated system we have
very accurately calculated the eigenstates and eigenvalues of the
$H_{\rm sys}$ checking that the tunneling time of our initial
state~(\ref{psi0}) is indeed very close to that estimated
by Eq. (\ref{initial}). This is illustrated in  Fig.~\ref{figure1}
where we show the time evolution of the probability of finding the particle
on the original well, for both the isolated and open system. Starting from
unit at $t=0$, the probability for the former decreases as the
particle tunnels through the barrier, reaching zero when $t\simeq \tau$.
For longer times (not shown in the figure) the particle tunnels back and
forth between the two wells, and as expected the probability is seen
to oscillate with a period close to $2\tau$.
 The behavior of the open system is in marked contrast with this. The probability
of remaining in the original well decreases but at a much slower rate when
compared to the open system. As we will see, this decrease is a consequence of
thermal activation rather than tunneling, which is suppressed at very early
times. The value of the probability never goes to zero, neither are any
oscillations observed. In fact the probability decreases monotonically  and for very long 
times we should expect it to approach $0.5$; when the system thermalizes 
it is equally probable to find the particle on each side of the barrier. 
On the same figure we also show the evolution of the linear entropy $S_L$
of the open system in terms
of the maximum of entropy allowed for the finite space representing
the Hilbert space of the system $\ln N$,
$S_L/\ln N=-\ln\left[ {\rm Tr}\rho^2\right]/\ln N$.
After some time the linear entropy reaches saturation 
suggesting that the dimension of the finite space ($N=40$) is too small. 
As a consequence the numerical results are less reliable after $t \sim \tau$.
Nevertheless it is clear from the plots that decoherence inhibits
tunneling well {\it before} this time.

\begin{figure}
\epsfxsize=7.6cm
\epsfbox{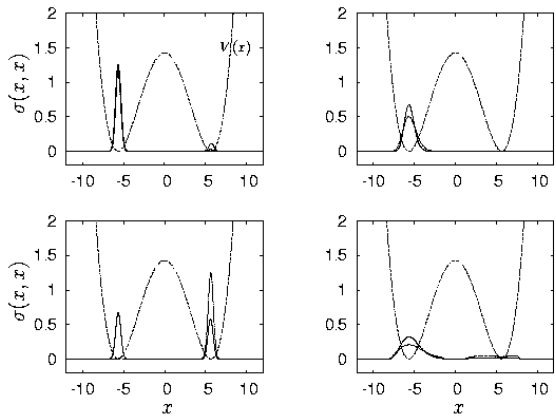}
\caption{Probability distribution $\sigma(x,x)$ for the isolated  (left)
and the open system (right) for $t=0$; $t=0.1 \tau$ and $t=0.2\tau$ (top)
and $t=0$; $t=0.5 \tau$ and $t=\tau$ (bottom).  As a reference, it is
drawn the shifted and scaled potential $V(x)$ in all the plots.
\label{figure2}}
\end{figure}

\begin{figure}
\epsfxsize=7.6cm
\epsfbox{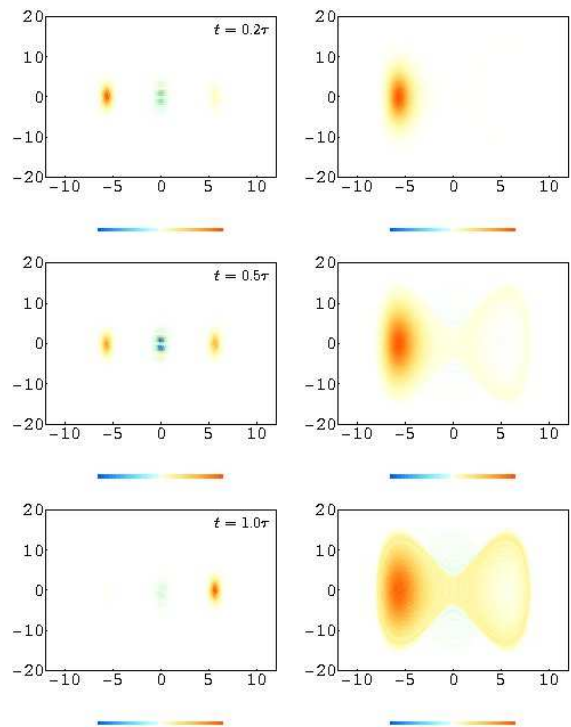}
\caption{Wigner distribution functions for the isolated (left) and open
(right) system, for the indicated times.
 Horizontal axis corresponds to $x$,
vertical axis to $p$. The medium grey shade on the background
corresponds to zero values for the Wigner function, lighter and
darker shades respectively to positive and negative values of $W(x,p)$.}
\label{figure3}
\end{figure}

The qualitative features of the evolution of both the isolated and the
open systems are illustrated
in Figs.~\ref{figure2} and \ref{figure3} where we can see, respectively,
the probability distribution
$\sigma(x,x) = \langle x | \rho_{\rm r} | x\rangle$ and the
Wigner function $W(x,p)$  for significant times.
Again, the contrast between their behaviour is very clear.
For very early times ($t\simeq 0.2\tau$), tunneling starts taking place
in the isolated system with $\sigma(x,x)$ becoming non-zero in the right-hand
well of the potential. The same effect can be observed in the Wigner function,
which also shows negative values in the centre of the phase space, indicating
clear quantum behaviour. For the same times, the open system shows no
signs of tunneling, with the 
particle  strictly confined to the original side of the potential.
The spread of both $\sigma(x,x)$ and $W(x,p)$ increases, as a consequence
of diffusion induced by the environment. As expected, since $t_D$ is very small
for this system, decoherence has clearly taken
place by this time and the Wigner function is strictly positive everywhere.
For $t=0.5 \tau$, both the probability distribution and the Wigner
function are symmetric for the isolated system. On the other hand, the wave
packet in the open system has continued to widen, and we see the first signs of
crossover above the barrier. As the tunneling time is reached,
though the system is still mainly localized on the original well,
it has become warmer and the Wigner function explores
a large region of phase-space, with thermal activation becoming significant.
Note that since areas of stronger non-linearity of the potential are now
occupied, one can observe slight negative valued fringes in the Wigner function.
This transitory behaviour is a well known consequence of the introduction of
non-linear effects in the system, and bares no relation with tunneling \cite{nos}.
At this time, on the other hand, the isolated system has fully tunneled and the wave packet
is centered around the right-hand minimum of the potential.   

\subsection{Thermal Activation in the Classical Limit}

In this section we will present a numerical example of classical 
high-$T$ thermal-activation. Our main goal is to confirm
that after decoherence takes place, a quantum system such as the 
one studied in Section \ref{numerical}, follows the behaviour of a purely
classical system, displaying thermal-activation.
 
 A classical statistical system is described by the Fokker-Planck equation
Eq.~(\ref{fp}). Here, instead of solving Eq.~(\ref{fp})
directly to obtain $W(x,p)$, we chose to evolve a very large
ensemble of classical particle trajectories interacting with a 
thermal bath via dissipation and noise terms. The equation of motion
for each particle is given by:
\begin{equation}
\ddot{x}(t)=-2 \gamma_0 \dot{x}(t) -V'(x(t)) + \xi(t),
\label{langevin}
\end{equation}
where $\xi$ is time-uncorrelated  Gaussian noise with variance
$\langle\xi(t)\xi(t')\rangle=\gamma_0 T \delta (t - t')$. It is 
straightforward to show that an ensemble of 
particles evolving according to the Langevin equation above, does indeed
obey the master equation Eq.~(\ref{fp}).
The numerical solution of a large number of equations of the type 
of Eq.~(\ref{langevin}) is trivial, 
offering an alternative to the direct solution of the
master equation as we have done so far. The initial conditions
are generated such that $x$ and $p$ are Gaussian random variables
distributed according to the classical analogue of the wave function
Eq.~(\ref{psi0}):
\begin{equation}
W_0(x,p)=\frac{1}{\pi} \exp\left[ -\frac{(x-x_0)^2}{2\sigma_x^2}
-2 \sigma_x^2 p^2\right].
\end{equation}
At arbitrary time $t$ we can obtain expectation values of physical
properties by averaging over the ensemble. The Wigner function
$W(x,p,t)$, can be determined by evaluating the fraction of particles
in the ensemble with position and momentum in the interval
$(x,x+dx)\times(p,p+dp)$. 

\begin{figure}
\epsfxsize=7.6cm
\epsfbox{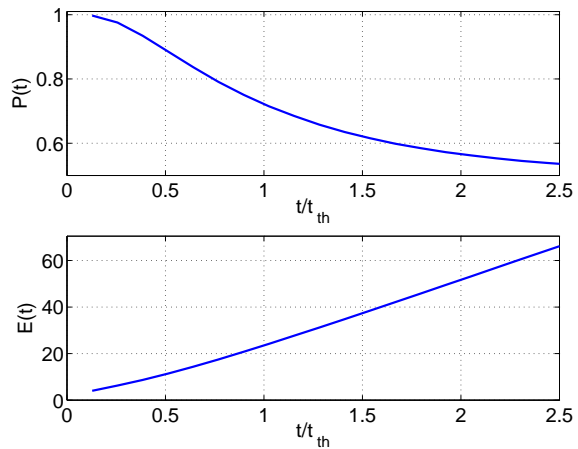}
\caption{Time dependence of the probability of remaining in the original side
of the potential (top) and average energy (bottom). Time is expressed in units
of the thermal-activation time scale Eq.~(\ref{t_th}).}
\label{class}
\end{figure}

\begin{figure}
\epsfxsize=8.6cm
\epsfbox{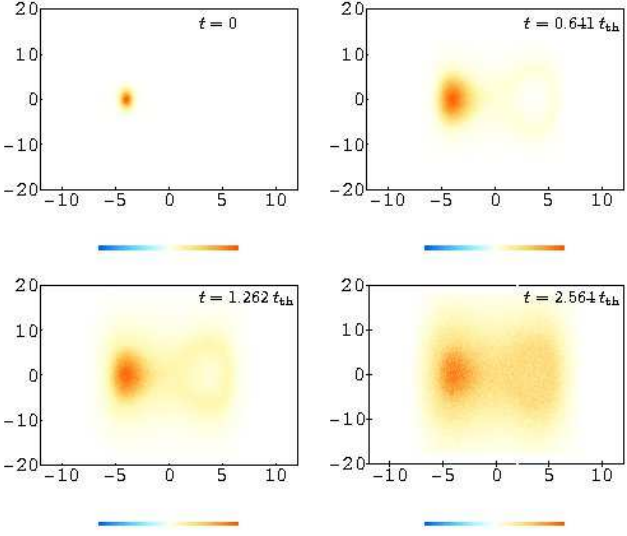}
\caption{Classical distribution function of the system for the indicated times.
Horizontal axis corresponds to $x$,
vertical axis to $p$. The medium grey shade on the background
corresponds to zero values for the Wigner function, lighter and
darker shades respectively to positive and negative values of $W(x,p)$}
\label{wigner_class}
\end{figure}

In Figs.~\ref{class} and \ref{wigner_class}
we show the results of a simulation with 
$\Omega^2=12$, $V_0=23$, $T=10^7$, $\gamma_0=2.5\times 10^{-9}$. Note that
choosing the set of  parameters used in Section \ref{numerical} would lead to
impractical simulation times. The qualitative aspects of the
dynamics of the two systems should be similar though, with the classical
simulation illustrating the generic properties of
the thermal-activation process.

 The thermal-activation time as estimated by Eq.~(\ref{t_th}) is given,
for this set of parameters, by $t_{\rm th}=390$. In Fig.~\ref{class}
we show both the probability of finding the particle in the left-hand
side of the potential and the mean energy of the system. As expected,
when $t\simeq t_{\rm th}$ the energy is of the order of the height
of the potential barrier. The probability at that time is $P\sim0.7$.
We simulated a series of similar processes with a wide range of parameters and
found that Eq.~(\ref{t_th}) holds very well over several
orders of magnitude of the quantities involved. In particular,
we found that the ``non-crossing'' probability at $t=t_{\rm th}$ is
always in the range $P\simeq 0.65-0.75$. The probability observed in
the quantum simulation of Section \ref{numerical} for $t_{\rm th}$ is
within this range. This  result should be taken qualitatively though,
since $t_{\rm th}$ is reached after entropy saturation has taken place.
The overall evolution of $P(t)$ in the classical case follows very closely
that for the quantum system after $t_{\rm D}$, with the probability decreasing
monotonically and approaching $0.5$ for large values of $t$.

 In Fig.~\ref{wigner_class} we have the phase-space probability distributions
(the classical Wigner function) for significant evolution times. Throughout the
evolution $W(x,p)>0$  as expected, since the Fokker-Planck equation
conserves the positivity of the distribution. As time progresses the initial
Gaussian packet widens, its energy increasing and allowing a larger fraction
of the ensemble's particles to explore further regions of phase phase. For
$t=t_{\rm th}$, when as defined, the average particle energy equals the potential
height, thermal crossing of the barrier starts to be significant. It is interesting
to note that for this period of evolution, the separatrix of the phase space shows
a high particle density on the right side of the potential. This confirms that the
particles crossing the barrier do so because their energy is of the order of the
barrier (corresponding to the separatrix energy). This obvious signature of classical 
thermal-activation  can also be observed in the quantum open system in
Fig.~\ref{figure3}. In the quantum isolated system on the other hand,
the Wigner function remains zero
in the separatrix region throughout the evolution. In this case,  tunneling can be
recognized by the large negative interference fringes in the origin of the phase-space.
Back in the classical system, we observe that for large $t$ dissipation and diffusion effects
combine to populate the central regions of the right-hand minimum of the potential.
Finally, the overall shape of the Wigner function becomes increasingly symmetrical,
with the system converging asymptotically to a thermal equilibrium state.

\section{Decoherence and tunneling at zero temperature}

At $T=0$ the time integrals in master equation~(\ref{master}) can be explicitly
calculated \cite{pau}.  We focus, as before, on Ohmic environments with spectral density
$I(\omega)= \frac{2}{\pi} \gamma_0 \omega \frac{\Lambda^2}{\Lambda^2+\omega^2}$.
After a rather lengthy calculation, the master equation at $T=0$ on the basis
of eigenstates of the isolated system can be written as
\begin{eqnarray}
\nonumber
\dot \rho_{\mu\nu} &=& - i \, \Delta_{\mu\nu} \rho_{\mu\nu} -\\
\nonumber
&-& \sum_{\alpha\beta}
\{ D_{\alpha\beta} x_{\mu\alpha} x_{\alpha\beta} \rho_{\beta\nu}-
   D_{\beta\nu} x_{\mu\alpha} x_{\beta\nu} \rho_{\alpha\beta}-\\
   \nonumber
  &-& D_{\mu\alpha} x_{\mu\alpha} x_{\beta\nu} \rho_{\alpha\beta}+
   D_{\alpha\beta} x_{\alpha\beta} x_{\beta\nu} \rho_{\mu\alpha}\} +\\
   \nonumber
&+& i \sum_{\alpha\beta}
\{ \gamma_{\alpha\beta} x_{\mu\alpha} x_{\alpha\beta} \rho_{\beta\nu}+
   \gamma_{\beta\nu} x_{\mu\alpha} x_{\beta\nu} \rho_{\alpha\beta}-\\
  &-& \gamma_{\mu\alpha} x_{\mu\alpha} x_{\beta\nu} \rho_{\alpha\beta}-
   \gamma_{\alpha\beta} x_{\alpha\beta} x_{\beta\nu} \rho_{\mu\alpha}\},
\label{masterT=0}
\end{eqnarray}
where the time dependent complex coefficients $D_{\alpha\beta} =D_{\alpha\beta}(t)$
and $\gamma_{\alpha\beta} =\gamma_{\alpha\beta}(t)$ are given by
\begin{eqnarray}
D_{\alpha\beta} &=& D(\Delta_{\alpha\beta}) + i\; \Delta_{\alpha\beta} \;f(\Delta_{\alpha\beta})\\
\gamma_{\alpha\beta} &=& -\frac{1}{2} \tilde\Omega^2(\Delta_{\alpha\beta}) - i 
\; \Delta_{\alpha\beta}\; \gamma(\Delta_{\alpha\beta})
\end{eqnarray}
with
\begin{eqnarray}
D(\Delta)&=&\frac{2\gamma_0}{\pi} \frac{\Lambda^2 \Delta}{\Delta^2 + \Lambda^2}\nonumber\\
&\times &\left[Shi(\Lambda t) \left(\frac{\Lambda}{\Delta}\cos\Delta t
\cosh\Lambda t +
\sin\Delta t \sinh\Lambda t\right)\right. \nonumber \\
&-& \left.
Chi(\Lambda t) \left(\frac{\Lambda}{\Delta}\cos\Delta t \sinh\Lambda t +
\sin\Delta t \cosh\Lambda t\right) \right. \nonumber \\
&+& \left. Si(\Delta t) \right]\,,\nonumber\\
f(\Delta)&=& 2\gamma_0 \frac{\Lambda^2}{\Delta^2 + \Lambda^2}\nonumber\\
&\times &\left[Shi(\Lambda t) \left(\frac{\Lambda}{\Delta}\sin\Delta t
\cosh\Lambda t -
\cos\Delta t \sinh\Lambda t\right)\right. \nonumber \\
&+& \left.
Chi(\Lambda t) \left(- \frac{\Lambda}{\Delta}\sin\Delta t \sinh\Lambda t +
\cos\Delta t \cosh\Lambda t\right) \right. \nonumber \\
&-& \left. Ci(\Delta t) - \ln\frac{\Lambda}{\Delta} \right].
\label{diff}
\end{eqnarray}
and
\begin{eqnarray}
{\tilde\Omega}^2(\Delta) &=& -\frac{2\gamma_0\Lambda^3}{\Lambda^2 +
\Delta^2} \left[1 - e^{-\Lambda t} \left(\cos\Delta t -
\frac{\Delta}{\Lambda} \sin\Delta t\right)\right]\,,\nonumber\\
\gamma(\Delta) &=&  \frac{\gamma_0\Lambda^2}{\Lambda^2 + \Delta^2}
\left[1 - e^{-\Lambda t} \left(\cos\Delta t + \frac{\Lambda}{\Delta}
\sin\Delta t\right)\right] \label{diss}
\end{eqnarray}
As before $\Delta_{\alpha \beta}=\omega_\alpha -\omega_\beta$, the frequency
difference between eigenstates $\alpha$ and $\beta$.
The set of coefficients $D_{\alpha\beta}$ encapsulates the effects of diffusion
 at $T=0$, with $D(\Delta)$ representing the normal diffusion and $f(\Delta)$ 
the anomalous one.
The others represent the effect of the environment through
the dissipation kernel $\eta$, with $\tilde\Omega(\Delta)$
the frequency shift and $\gamma(\Delta)$ the dissipation coefficient.
The last two reach constant asymptotic values for $\Lambda t \gg 1$.

Even though the time dependent functions (\ref{diff})
reach an asymptotic constant value for $\Delta_{\alpha\beta} t \gg 1$
and $\Lambda t \gg 1$,
for the problem we are going to analyze we will never reach
a regime where all the coefficients involved by equation (\ref{masterT=0})
are constant.
That is, the expressions in (\ref{diff}) are constant for
{\it all} $\alpha, \beta$ when $t \gg 1/\Lambda$ and
$t \gg 1/\Delta_{1,0} \approx \tau/3$,
due to Eq. (\ref{initial}).
As the functions $Si(t)$ and $Ci(t)$ converge toward its asymptotic values only
very slowly, we will never reach this regime.

As for the high-T limit we want to estimate the decoherence time scale.
For this purpose we will analyze the decoherence process in a simple case:
$\Psi(x,t=0) = \Psi_1(x) + \Psi_2(x)$,
where
\begin{equation}
\label{psi12}
\Psi_{1,2} = N \exp\left(-\frac{(x\mp L_0)^2}{2\delta^2}\right)
\exp(\pm i P_0x),
\end{equation}
with $N$ a normalization constant, and $\delta$ the initial width of the
wave packet.

As it was defined in the previous literature (see for example
\cite{Paz_Zurek,phz}), the effect of decoherence is produced by an
exponential factor $\exp(-A_{\rm int})$, defined as

\begin{equation}
\exp(-A_{\rm int}) = \frac{1}{2}\frac{W_{\rm int}(x,p)|_{\rm
peak}}{\left[W_1(x,p)|_{\rm peak} W_2(x,p)|_{\rm
peak}\right]^{\frac{1}{2}}},\end{equation}
where $W{\rm int}$ is the Wigner's interference term, coming from the superposition 
of the two states $\Psi_{1,2}$. 

In a very crude approximation one may drop all nonlinear
terms on the Hamiltonian of the system and then estimate the
decoherence time-scale from (see Ref.\cite{pau} for details)
\begin{equation}
\dot{A}_{\rm int}\approx 4 L_0^2 D(\Delta) - 2 f(\Delta)\,,
\label{aint}
\end{equation}
where $L_0$ is the spread of the state. 
In order to evaluate the decoherence time $t_{\rm D}$, we have to solve
$1\approx A_{\rm int}(t = t_{\rm D})$. From Eq. (\ref{aint}) it is not possible
to find a global decoherence time-scale at $T = 0$. Nevertheless,
we can find limits in which we are able to give different scales
for decoherence.

When $\Delta t \ll 1$
(for times $\frac{1}{\Lambda} < t < \frac{1}{\Delta}$), we can
approximate $A_{\rm int}$ using the asymptotic limits of $Si$ and
$Ci$ by
\begin{equation}
A_{\rm int} \approx \frac{8\Lambda^2}{\Lambda^2 + \Delta^2} \gamma_0 \left[
\frac{L_0^2}{2\pi} (\Delta t)^2 + t ~ (\log\Lambda t + \Gamma - 1)\right]  ,
\label{a2}
\end{equation}
resulting in a decoherence time bound
\begin{equation}
\label{tDlarge}
t_D \leq \frac{1}{8  \gamma_0}.
\end{equation}

For large frequency $\Delta$, such
as $\Delta \sim \Lambda$, it is easy to see that

\begin{equation} A_{\rm int} \sim 2 L_0^2 \gamma_0 \Lambda t +
 4 \gamma_0 \left(t ~ Ci(\Lambda t) -
\frac{\sin \Lambda t}{\Lambda}\right),
\end{equation} giving a very short decoherence time-scale,

\begin{equation}
t_D \sim \frac{1}{2 L_0^2 \gamma_0 \Lambda}.
\label{td1}\end{equation}
This result will be valid as long as the product $L_0^2 \gamma_0
\leq 1$, allowing us to neglect the initial transient.

As this decoherence time scale was derived after dropping all
nonlinear terms on the Hamiltonian of the system, it is then valid only for
linear systems. We can only
expect it to be of some use when we begin with a narrow initial
state located
at one of the potential minima of the system because for a while it will
evolve as in an harmonic oscillator potential.
After some (short) time
the nonlinearities will generate interferences dynamically \cite{nos}.
Then, we should expect equation (\ref{tDlarge}) to be only accurate
if decoherence happens early enough, before non-linear effects kick in.

\subsection{Numerical Results}

We have numerically solved equation~(\ref{masterT=0}) using a
standard adaptative step-size fifth order
Runge-Kutta method for different parameters of the  system and
the environment.
All results were found to be robust under changes on the parameters
of the integration method.

As an example we have chosen $\Omega=100$ and $V_0 = 200$ for the system
for which the estimated tunneling time scale is $\tau\approx 158.27$.  As for the
high-T limit, we desire decoherence to occur before tunneling,
so we have set the parameters of the environment according to
$a t_{\rm D} = \tau$ with $a=10$.  Then from equation (\ref{tDlarge})
$\gamma_0 = a/(8 \tau) \approx
0.007897$.  We set the frequency cutoff to $\Lambda =10 V_0= 2000$. With
 this set of parameters, and 
taking into account Eq.(\ref{diss}), we see that the effects of the
frequency shift in the initial state can be neglected. In fact, 
it is easy to check that for these values, ${\tilde \Omega}^2$ is $0.32\%$ of 
$\Omega^2$. Therefore we can safely neglect the error induced by taking the
initial state to be given by the vacuum of an harmonic oscillator
of frequency $\Omega$, rather than ${\tilde \Omega}$.

\begin{figure}
\epsfxsize=7.6cm
\epsfbox{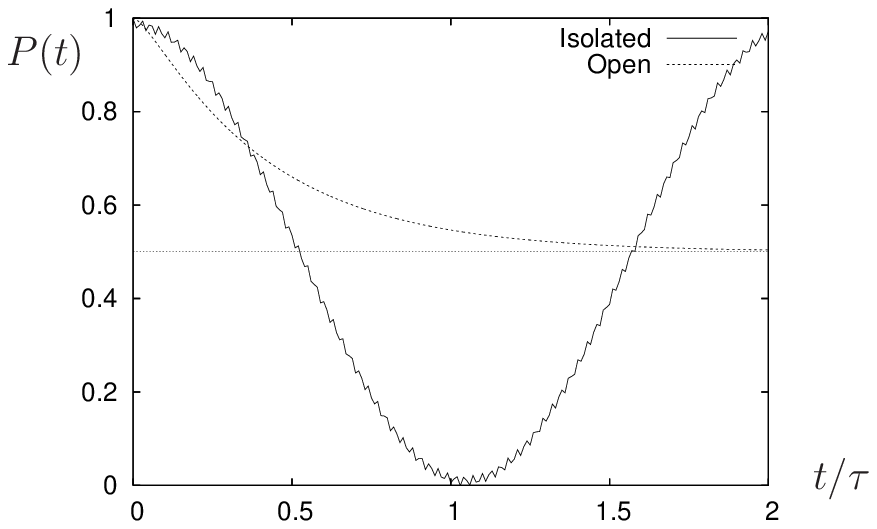}
\caption{Time behavior of the probability to stay on the left of the
barrier. Time is measured in units of the estimated tunneling time $\tau$.}
\label{fig1T=0}
\end{figure}

\begin{figure}
\epsfxsize=7.6cm
\epsfbox{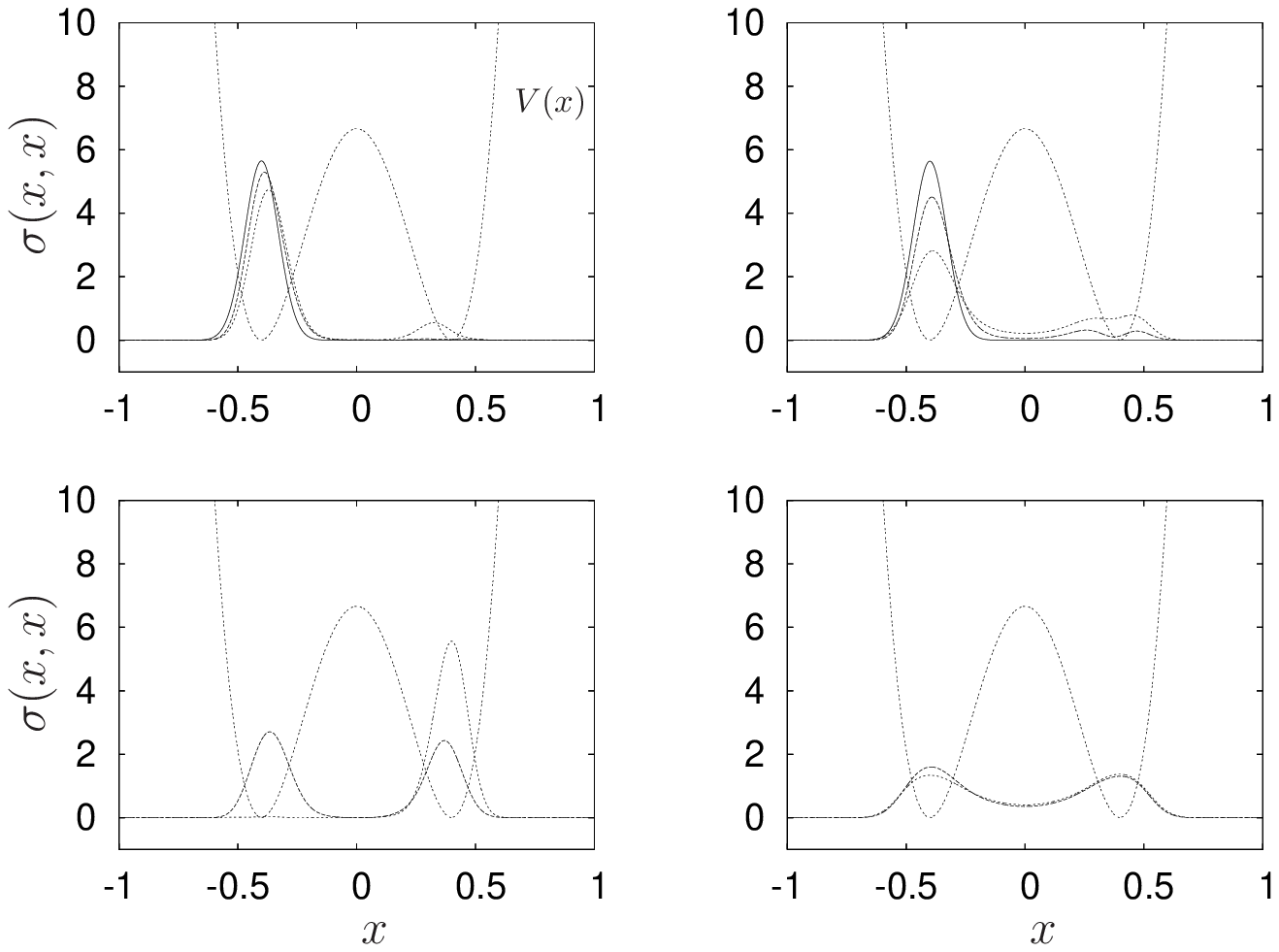}
\caption{Probability distribution $\sigma(x,x)$ for the isolated (left)
and the open system (right) for $t=0$; $t=0.1\;\tau$ and $t=0.2\;\tau$ (top)
and $t=0.5\;\tau$ and $t=\tau$ (bottom).  As a helpful reference, the
scaled potential $V(x)$ is drawn in all the plots.
\label{fig2T=0}}
\end{figure}

\begin{figure}
\epsfxsize=7.6cm
\epsfbox{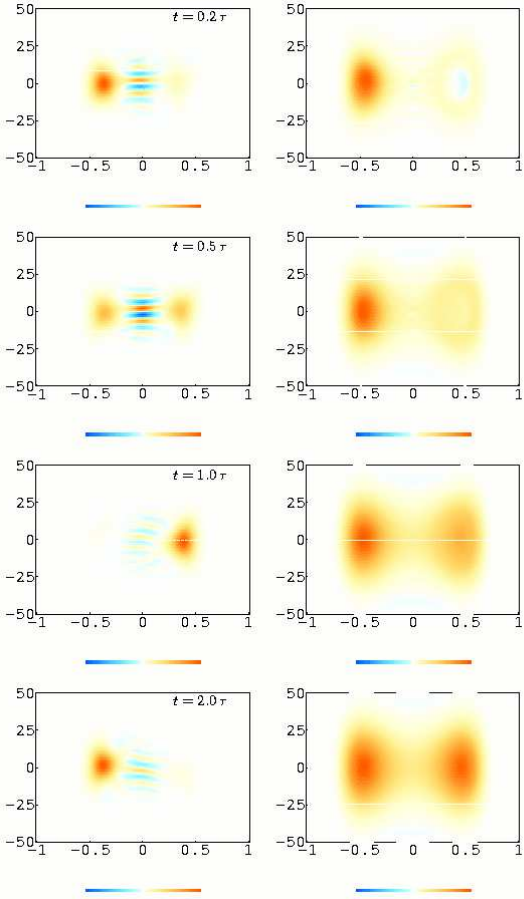}
\caption{Wigner distribution functions for the isolated (left) and open
(right) system, for the indicated times. Axes and gray shades are similar to 
the ones defined in Fig.~\ref{figure3}.}\label{fig3T=0}
\end{figure}

Fig.~\ref{fig1T=0} shows the probability of staying in the original well,
$P(t) = \int_{-\infty}^0 \; dx\;\sigma(x,x)$ in terms of the time measured
in units of the estimated tunneling time $\tau$, while
Fig.~\ref{fig2T=0} and \ref{fig3T=0} show the probability distribution $\sigma(x,x)$
and the Wigner function of the system, respectively,
 for the indicated
times, both for the isolated and the open system.

 The evolution of the open system at zero temperature shows  similarities
but also relevant differences
with the analogous situation in the high-T limit. For very 
early times the probability of staying on the initial side of the potential
decreases quickly, approaching $0.5$ as soon as $t\sim 2\tau$. There are no
signs of the particle tunneling back, as expected, since $t_D$ is taken to
be smaller than the tunneling time-scale. Also, the asymptotic behaviour of
 $P(t)$ shows
no oscillatory behaviour as would be expected if tunneling played any role in 
the late-time dynamics. Instead we see what looks like a quick ``equilibration'' into
a state where the particle is equally likely to be on either side of
the potential barrier. Both the probability distribution plots and
the Wigner functions corroborate this picture. From very early-times, 
the negative regions of $W(x,p)$ in the $T=0$ case are considerably 
suppressed when compared with the closed system, suggesting that tunneling
has a small contribution to the evolution. For $t>\tau/2$,
 $W(x,p)$ becomes positive definite and the system displays no tunneling
 oscillations. As in the high-T case, the separatrix
becomes densely populated when cross-over starts being significant. Both this
and the fact that for late times $\sigma(x,x)$ and $W(x,p)$ are symmetrical
around $x=0$, suggests we should be able to describe the dynamics of the
open system in terms akin to classical activation.
 
 When trying to interpret the post-decoherence behaviour of the open system,
several features of its dynamics should be kept in mind. Firstly, one should
emphasize that the initial condition is clearly not the ground state of the 
total action Eq.~(\ref{action}). As soon as the interaction between the
main system and the environment is turned on, at $t=0$, the system will
find itself in an excited state. In relation to the new minima of the 
potential, the environment will have a non-zero amount of energy. 
From a purely classical point of view, this energy
cannot be responsible for the excitation of the particle over the potential barrier.
In fact, the height of the potential increases in relation to the new vacuum, in a 
way such that the {\it total} energy of the full system is still lower than the
barrier separating regions of positive and negative $x$. This argument can
be made more quantitative in the following way:
the full potential for the system plus environment
is $ V(x,q_n) = V_{\rm sys}(x) + V_{\rm env}(q_n) + V_{\rm int}(x,q_n)$
with
$V_{\rm sys}(x) = -1/4 \Omega^2 x^2 + \lambda x^4$, 
$V_{\rm env}(q_n)= \sum_n 1/2 \omega^2 m_n^2 q_n^2$, 
and $V_{\rm int}(x,q_n)= \sum_n C_n x q_n$. Classically, the initial 
 condition is $ x=-\Omega/\sqrt(8 \lambda )$,
 and, because the environment is at T=0, $q_n=0$. So, for the
 full action, the energy terms of the initial condition are
 given by $V_{\rm sys} = - \Omega^4/(64 \lambda)=V_0$ 
(the minimum of $V_{\rm sys}$), $V_{\rm env}=0$, and  $V_{\rm int}=0$;
and so $V=V_0$. Note that classically, the value of the total energy is the same as the
 energy of the isolated main particle, even when the interaction
 with the environment is ``switched on". This is a consequence
 of taking zero temperature for the environment. It is true that
 when the full system is considered, we are no longer in the state
 of minimum of energy, $V_0$ corresponding to an excited state. Nevertheless,
 this initial energy cannot be responsible for making the particle
 cross the potential barrier. The classical trajectory of a
 particle going over the barrier would have necessarily $x=0$ at some
 point. If $x=0$ the value for the total energy
 must be positive $V > 0$. Since for the initial state implies $V_0<0$, this can
 never happen. In other words, when the interaction is switched on, the system
 does ``gain" energy relatively to the new vacuum, but the height of the
 barrier increases by the same amount, so classical activation
 cannot take place.

 Note that the fact that there are no fluctuations in the environment
 classically at T=0, plays a crucial role in this reasoning. Even
 for small but finite T, the energy of the environment would
 go as T. By choosing T small enough, this contribution could always
 be made smaller than the barrier height. As a consequence, and in contrast 
with the high-T case, we will not
be able to describe the evolution of the quantum system after classicalization
by simply taking its classical exact equivalent. The
quantum fluctuations present in the initial state of the environment must
play a role in the ``activation''. One should note that these fluctuations are
not ``vacuum fluctuations" of the full system. Nevertheless, the
fact that they have such a clear effect on the evolution of the system is quite
remarkable. Whereas in the high-T case the quantum and classical oscillators
composing the bath had identical distributions, they behave in a markedly different way
as $T\rightarrow 0$. The quantum nature of the environment,
which could be ignored at high-T, leads in this limit to important non-negligible
effects. 

In terms of the master equation, the quantum fluctuations of the bath
oscillators generate non-zero $f(t)$ and $D(t)$ terms, 
making diffusive phenomena possible. This is particularly true of the anomalous
diffusion coefficient $f(t)$ that depends logarithmically on the cutoff
$\Lambda$ and thus can be considerably large \cite{pau}. Diffusion effects induced
by quantum fluctuations  are thus responsible for exciting the particle
over the potential barrier. Though this process is very different from high-T
thermal activation, we conjecture that it may still be interpreted in terms of a
modified classical setting. The key ingredient is that the classical bath
 should mimic the
properties of the quantum $T=0$ environment. Considering the classical and
quantum versions of the noise kernel $\nu(s)$,
it is possible to show that a bath of classical oscillators with a frequency
dependent temperature $T(\omega)=\hbar \omega/2$ should reproduce the 
effects of the initial quantum state. In fact, for this choice of classical environment
one obtains $f(t)$ and $D(t)$ terms identical to those of the $T=0$ quantum case.
Our main point is that after decoherence takes place, a quantum open
system at $T=0$ should behave as a classical open system in contact with a
classical bath whose oscillators are excited in a way that
reproduces the fluctuations of the corresponding quantum environment.
In order to fully understand this correspondence, one should simulate a classical system
interacting with this type of generalized bath, reproducing the results 
of the quantum case and obtaining the same time-scales for fluctuation-induced
activation \cite{calzetta1,calzetta2}. We will leave a detailed study of this 
type of system to a future publication \cite{future}.


\begin{figure}
\epsfxsize=7.6cm
\epsfbox{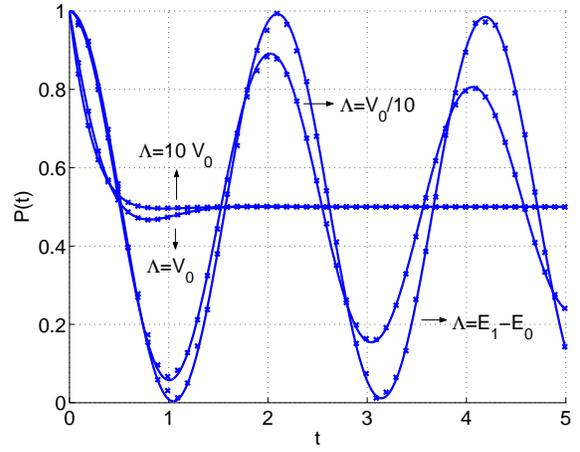}
\caption{Probability to stay on the original well for different 
values of $\Lambda$ (other parameters are fixed as in previous plots). 
The crosses are a subset of the simulation data (not all data points 
are shown so that the fit curves can be visible). The solid lines 
correspond to non-linear chi-squared fits of the data to the
expression in Eq.~(\ref{fit}).
Time is measured in units of the closed system tunneling time $\tau$.}
\label{prob-L2}
\end{figure}

\begin{figure}
\epsfxsize=7.6cm
\epsfbox{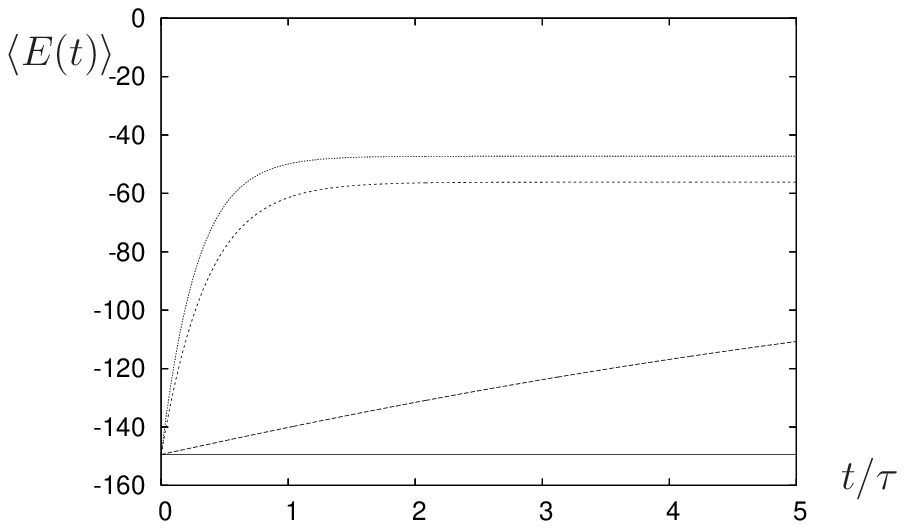}
\caption{Evolution in time of the mean energy of the main system for the 
same set of parameters as in Fig. \ref{prob-L2} (time in units of the 
tunneling time $\tau$). Solid line correspond to the 
smallest value of $\Lambda$; straight dashed line to $\Lambda = V_0/10$. 
Dotted lines are larger values of the cutoff: $\Lambda = 10 V_0$ on top, 
and $\Lambda = V_0$ below.}
\label{energia-L2}
\end{figure}

\begin{figure}
\epsfxsize=7.6cm
\epsfbox{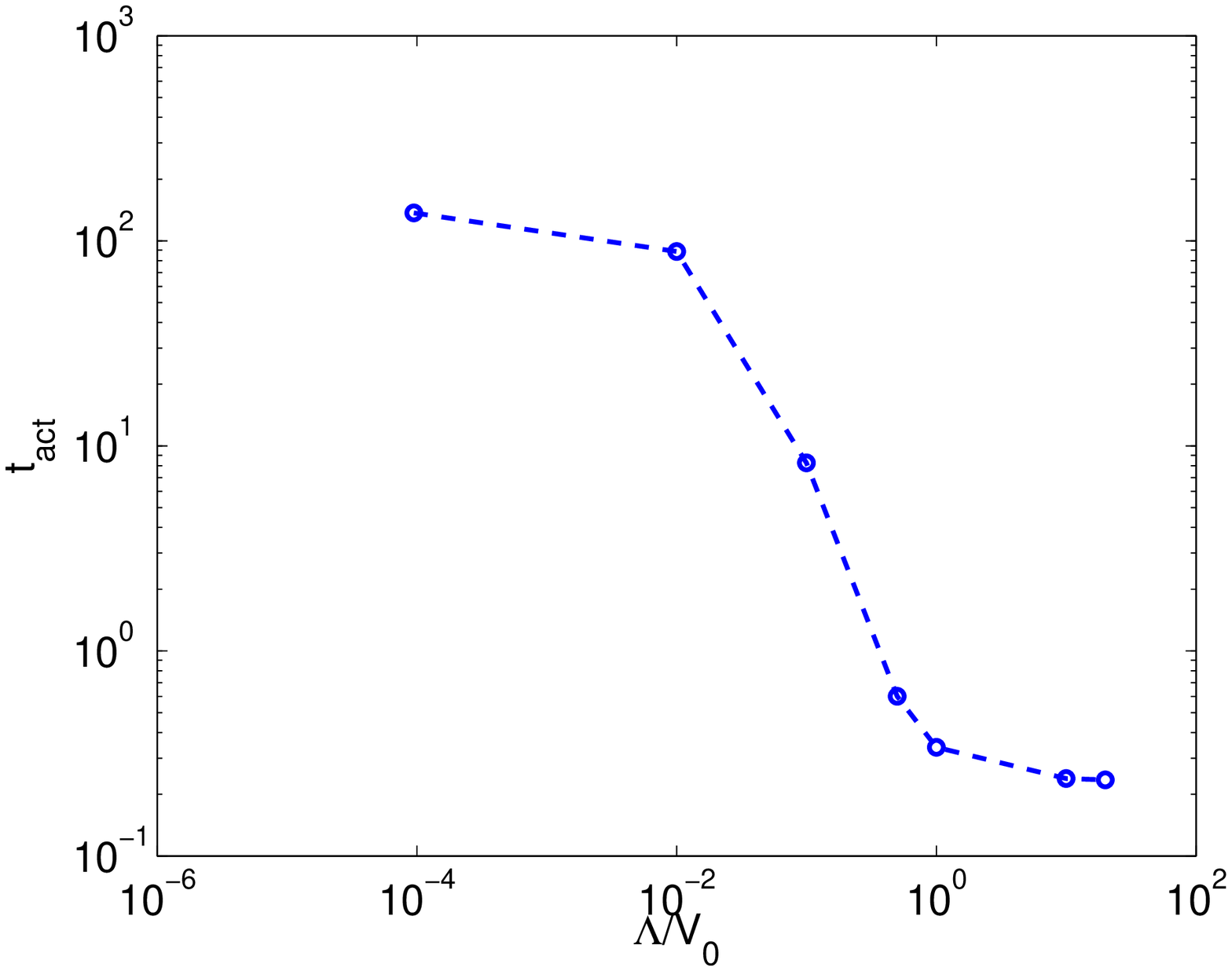}
\caption{Log-log plot of the activation time $t_{\rm act}$ as a
function of $\Lambda/V_0$.}
\label{t_act}
\end{figure}

 A second question concerns the interplay of decoherence and excitation processes in
the $T=0$ case. For both quantities, the value of the environment frequency
cutoff $\Lambda$, seems to play an important role, affecting both the decoherence
time, and the excitation process in the same direction. Since we do not have
explicit estimates of the ``activation'' time in terms of $\Lambda$, it is hard to
predict whether there is a regime for which decoherence happens fast enough and
excitation is considerably delayed. Numerical results presented in
Figs. \ref{prob-L2}
and \ref{energia-L2} suggest that this is not possible. The two figures
show respectively the probability to stay on the original well and the energy of 
the main system for several choices of the cutoff. $\Lambda$ varies
from  the smallest frequency present in the system; i.e. the difference between the first
 exited and the ground state energy levels, $E_1 - E_0$; and
 $\Lambda = 10 V_0$. Also shown are two intermediate cutoff values, $\Lambda = V_0/10$ and
 $\Lambda = V_0$. 
     By lowering $\Lambda$, 
the ``activation'' time is indeed
postponed, but so is decoherence. In this situation the particle is simply
able to tunnel back and forth the two minima for a longer period. Higher values
of the cutoff, on the other hand lead to both fast decoherence and fast
``activation''. As a result we were never able to localize the particle
on one of the wells, with tunneling and ``activation'' being simultaneously
suppressed. 
     
  The dependence of the activation time on the environmental cutoff
  frequency can be made more quantitative by fitting the probability
 for the particle to remain in the original well to a simple evolution
  expression.
  In Fig.~\ref{prob-L2}  a selection of simulation points 
  (crosses) is shown against a fit (solid curves) of the form:
  \begin{equation}
     P(t)=\frac{1}{2}+\frac{1}{2} \cos(\pi t/\tau) \exp(-t/t_{\rm
     act}).
  \label{fit}
  \end{equation}
  The analytical expression fits the data extremely well,
  allowing us to determine for each choice of $\Lambda$ the two 
  relevant time-scales, $\tau$ and $t_{\rm act}$. In Fig.~\ref{t_act}
  the activation time measured in this way is shown as a function
  of the cutoff parameter. This figure includes results 
  for a larger number or curves, spanning several orders of magnitude
  of $\Lambda$. As expected the activation time decreases initially
  as the value of the cutoff increases. A change of regime is reached
  when $\Lambda$ is of the order of magnitude of the height of the
  potential barrier $V_0$. For all values of $\Lambda<V_0$, tunneling
  is still observed, and indeed, the tunneling time, as measured by
  $\tau$ obtained from the fit, deviates very little - less than
  $5\%$ - from the value for the isolated system. For higher values 
  of $\Lambda$ tunneling is completely suppressed, with the oscillating
  term in (\ref{fit}) becoming irrelevant for the fit. 
  This suggests that $\Lambda\simeq V_0$ can be taken as the threshold 
  for the fluctuations to play the principal role in the evolution, 
  with excitation becoming the dominant processes in the dynamics in
  this regime.
  In all cases, the long time limit value for the probability seems to 
  be  $0.5$ to a very good accuracy. As an extra check we re-fitted 
  the data allowing the asymptotic value of $P(t)$ as an extra free parameter.
  In the whole range of $\Lambda$ studied, the final probability always
  differed by less than $0.8\%$ from $0.5$. The values for $t_{\rm act}$
  obtained in the fit with the extra parameter changed by less than $6\%$
  when compared with the results shown in Fig.~\ref{t_act}.

 Similar properties can be observed in terms of the 
 energy of the system in Fig. \ref{energia-L2}, where we plot the mean energy
 of the (main) system as a function of time, 
 for the same set of parameters used in Fig. \ref{prob-L2}. Clearly, the ``activation''
 process for high frequency cutoff is accompanied by a fast increase
in the energy of the system. Once again, the fact that the energy of the main system is 
considerably lower than the barrier height for low $\Lambda$, supports the interpretation 
that quantum fluctuations are behind the excitation mechanism.

 \begin{figure}
\epsfxsize=7.6cm
\epsfbox{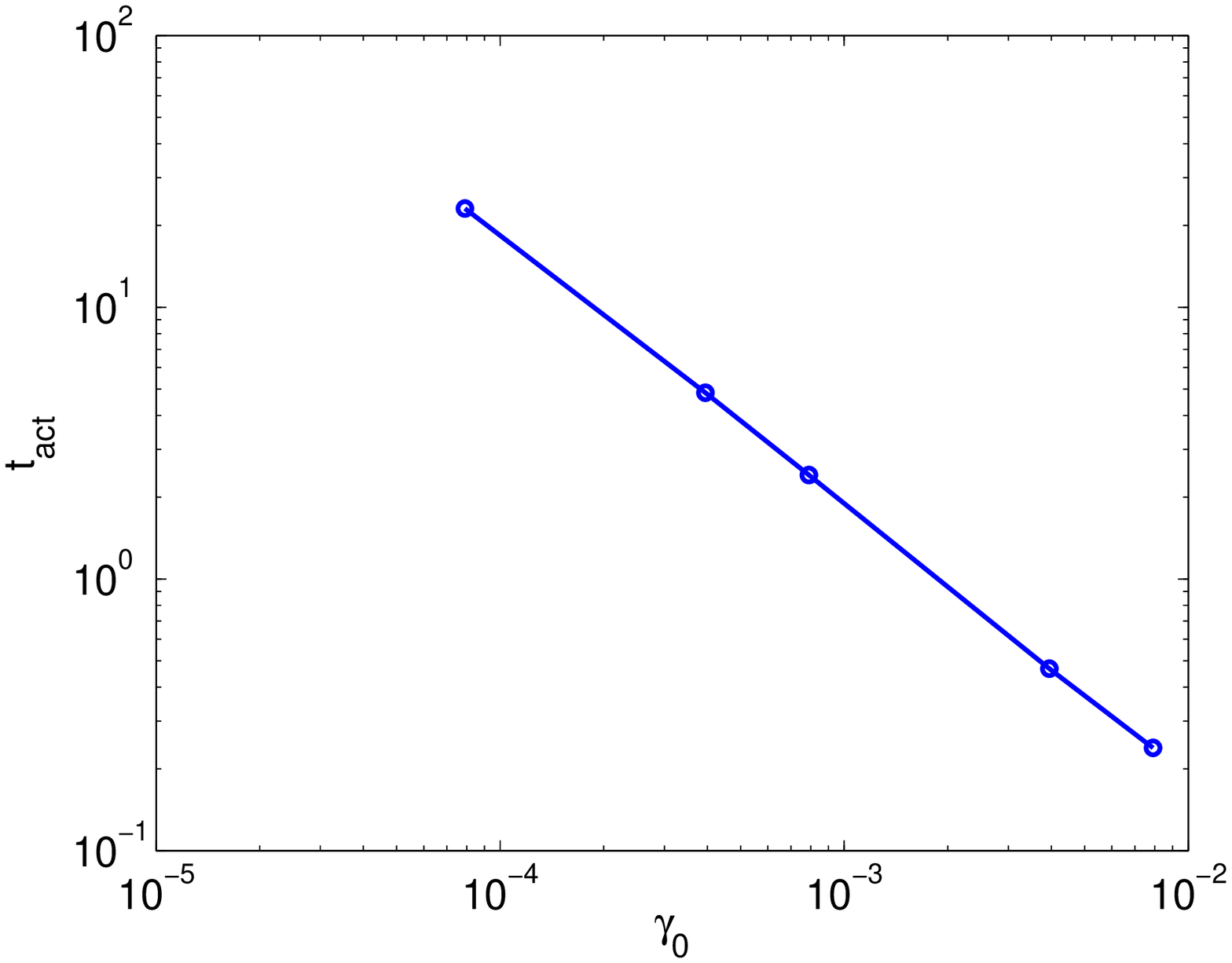}
\caption{Log-log plot of the activation time $t_{\rm act}$ as a
function of $\gamma_0$ for $\Lambda=10 V_0$.}
\label{t_act_gamma}
\end{figure}

 Finally, we looked at how the value of $\gamma_0$ affects the
 overall pattern of evolution, in the case of a cutoff
 $\Lambda = 10 V_0$. We found that as $\gamma_0$
 decreases, as expected, tunneling reappears and the activation time
 increases as shown in Fig~\ref{t_act_gamma}. Interestingly 
 the value of the measured tunneling time $\tau$ varies with
 $\gamma_0$. For very small values of the coupling, we obtain $\tau\sim 1$ in
 units of tunneling time. This value increases with $\gamma_0$ up to $50\%$
 of the original tunneling time. At this point, tunneling is suppressed and, as in the
 case discussed above, the oscillatory term in the fit can be
 ignored. This suggests renormalization effects are likely to play
 a role in this case,  as the strength of the interaction with the
 environment becomes larger. Also worth remarking is the fact that
 within the region of parameters tested, $t_{\rm act}$ seems to vary
 as the inverse of $\gamma_0$. This is reminiscent of Eqs.~(\ref{tDlarge}) and
 (\ref{td1}) indicating once again that decoherence and activation
 for this
 type of system are closely related. Overall, a very rich
 structure seems to emerge from the interplay of several physical
 mechanisms taking place simultaneously at $T=0$. A detailed numerical
 study of these will be the focus of a future publication.

\section{Final Remarks}

We have analyzed a simple time-independent bistable system,
by following the quantum evolution of a particle initially
localized at one of the minima of the potential, when coupled to an external
environment at both zero and high temperatures.
When isolated, the particle undergoes tunneling through the barrier.
For the open system, we described the effects of dissipation and diffusion on
its dynamics in terms of three main phenomena:
decoherence, tunneling and thermal activation. We estimated the corresponding 
time-scales analytically and showed that, depending on the parameters of
the system and its environment, these processes can be made to act 
independently on the evolution. The numerical results
confirmed the analytical estimates and allowed us to illustrate the 
distinct properties of the three types of process involved in the evolution. 

 For the closed system, the numerical simulations displayed all the expected
features of standard quantum tunneling, with the wave packet bouncing back and
forth between the two vacua, with a rate given by the estimated
tunneling time $\tau$. In the high-T regime, for an appropriate region of
parameter space, the open system was shown to evolve in a fundamentally different 
way, with the probability for the particle to be found in the original well
decaying at a much slower rate. With basis on the relevant time scales determined 
analytically, we explained the absence of tunneling in this case as a consequence 
of early time classicalization at $t_D$. The later time evolution on the other
hand, was interpreted as a result of classical thermal activation. This picture
was confirmed by looking at the evolution in terms of the space-probability 
distribution and the Wigner function. In contrast with the tunneling dynamics, 
the Wigner function for the open system became strictly positive soon after
the decoherence time, evolving classically for most of the simulation time. 
For late times, the properties of the Wigner function were characteristic
of thermal activation, with the phase-space separatrix becoming heavily populated
as the particle crossed over the potential barrier.
As an extra check, we evolved a similar classical system and confirmed that the
concentration of probability around the separatrix at $t_{\rm th}$ does signal the onset
of thermal activation. 
It is worth mentioning that the
environment can, for both high and zero temperatures, be tailored so as to have
tunneling before decoherence
or in fact any other permutation of the three processes describing the dynamics.
This can be easily seen  by comparing the
estimated time-scales for each process. Though not shown here, we have
checked  numerically that all these cases are indeed possible.

The evolution of the open system at zero temperature shows subtly different
and, in some ways, unexpected properties. Tunneling is also
undoubtedly quickly suppressed, as can be seen by inspecting either the
probability of the particle to remain on the original well or the evolution of
its Wigner function. In both cases we observe typical classical features
since very early times. Nevertheless, at $T=0$, the quantum fluctuations of 
the environmental oscillators, absent in a
purely classical evolution, lead to non-zero diffusive terms.
Their effect is felt primarily through the anomalous diffusion coefficient $f(t)$
that can have a large magnitude. We conjecture that these non-trivial
diffusion effects induced by the quantum environment are large enough to
excite the particle over the potential barrier. This is to be contrasted 
with the case where the environment is classical forbidding any kind of activation phenomena.
Though the late time evolution in the presence of a quantum vacuum
is in nature very different from high-T thermal activation, we suggest 
that it could still
be interpreted in terms of a purely classical setting, if the environment
oscillators are described by a particular non-thermal statistical state.
We will pursue this line of enquire in depth in a forthcoming publication.

\section{Appendix}
\label{apA}

Written on the basis of eigenstates $|\mu\rangle$ of the isolated
system $H_{\rm sys} = \frac{p^2}{2} + V(x)$, Eq. (\ref{master}) reads
\begin{eqnarray}
\dot \rho_{\mu\nu} = - \imath \, \Delta_{\mu\nu} \rho_{\mu\nu} -
&\sum_{\alpha\beta\gamma} \rho_{\alpha\beta} \left[  \int_0^t d\tau \right.\,
\nu(\tau)  A_{\alpha\beta\gamma}^{\mu\nu}(\tau)  & -\nonumber \\
&-  i \left. \int_0^t  d\tau \, \eta(\tau)
B_{\alpha\beta\gamma}^{\mu\nu}(\tau) \, \right]\,. &  \nonumber
\end{eqnarray}
The time-dependent coefficients $A$ and $B$ are sums of four terms of the
form $x_{\mu\alpha} x_{\alpha\beta} {\rm e}^{\imath \Delta_{\gamma\alpha} \tau}$:
\begin{eqnarray}
\nonumber
\left.\begin{array}{l}
A_{\alpha\beta\gamma}^{\mu\nu}(\tau)\\
\\
B_{\alpha\beta\gamma}^{\mu\nu}(\tau)
\end{array} \right\}
&=&
x_{\mu\gamma} x_{\gamma\alpha} \delta_{\beta\nu} {\rm e}^{-\imath \Delta_{\gamma\alpha} }\mp \\
&& \mp x_{\mu\alpha} x_{\beta\nu} {\rm e}^{-\imath \Delta_{\beta\nu}} -\nonumber\\
&& - x_{\mu\alpha} x_{\beta\nu} {\rm e}^{-\imath \Delta_{\mu\alpha}} \pm \nonumber\\
&& \pm x_{\beta\gamma} x_{\gamma\nu} \delta_{\alpha\mu} {\rm e}^{-\imath \Delta_{\beta\gamma}}.
\nonumber
\end{eqnarray}
Thus, all the time integrals appearing on the master equation have the form:
\begin{eqnarray}
\nonumber
\int_0^t d \tau \, \nu(\tau) \, {\rm e}^{\imath \Delta_{\alpha,\beta} \tau}  \;\;\; {\rm or} \;\;\;
\int_0^t d \tau \, \eta(\tau) \, {\rm e}^{\imath \Delta_{\alpha,\beta} \tau} \nonumber.
\end{eqnarray}
These integrals can be explicitly calculated only if the spectral density of the environment
is specified.  We have supposed an Ohmic environment, for which
$I(\omega) = 2 \gamma_0 \frac{\omega}{\pi} \frac{\Lambda^2}{\Lambda^2+\omega^2}$
where $\Lambda$ represents a high-frequency cutoff and $\gamma_0$ is a constant
characterizing the strength of the interaction with the environment (we have set the mass equal 
to one). 
For this environment the temperature-independent $\eta$-integrals can be easily calculated to be:
\begin{eqnarray}
\nonumber
\int_0^t d\tau \eta(\tau) {\rm e}^{\imath \Delta \tau} =
\tilde\Omega^2(\Delta,t) + \imath \Delta \gamma(\Delta,t),
\end{eqnarray}
where
\begin{eqnarray}
\tilde\Omega^2(\Delta,t)&=&
- \frac{2\gamma_0 \Lambda^3}{\Lambda^2+\Delta^2}
\left[1-{\rm e}^{-\Lambda t} (\cos\Delta t -
\frac{\Delta}{\Lambda} \sin\Delta t ) \right], \nonumber \\
\gamma(\Delta,t)&=&
\frac{ \gamma_0\Lambda^2}{\Lambda^2+\Delta^2}
\left[1-{\rm e}^{-\Lambda t} (\cos\Delta t + \frac{\Lambda}{\Delta}
\sin\Delta t ) \right],  \nonumber
\end{eqnarray}
are the frequency-shift and dissipation coefficients, respectively.

The $\nu$-integrals are not so easily calculated and so here we resort to a
Markovian approximation.  We will assume $\Lambda \gg \Delta_{\alpha\beta} \;  
\forall \; \alpha, \beta$.
Thus, the kernels are strongly peaked around $t=\tau$, and  the environment has a very short
correlation time.  Therefore the integrals can be extended over the entire interval
$[0,\infty)$.  If we further assume that the temperature is
very high, that is $T \gg \Delta_{\alpha\beta} \;  \forall \; \alpha, \beta$, the
$\nu$-kernel is reduced to a delta distribution function, and the
time integrals are simply \footnote{Actually this last assumption accounts for the former one, for
the high-temperature limit reduces to a Markovian approximation.}
\begin{equation}
\nonumber
\frac{\pi}{2} I(\Delta) \coth(\frac{\beta \Delta}{2}) \approx 2 \gamma_0 T = D
\end{equation}
Finally, after some algebraic manipulations, the master equation reads as
Eq. (\ref{rhomunu})
\begin{equation}
\dot \rho_{\mu\nu} = - \sum_{\alpha\beta} M_{\mu\nu\alpha\beta} \;
 \rho_{\alpha\beta}\,,
\nonumber
\end{equation}
where the {\it time-independent} coefficient $M$ is
\begin{eqnarray}
M_{\mu\nu\alpha\beta} &=& \imath \delta_{\alpha\mu} \delta_{\beta\nu} \Delta_{\alpha\beta}
+ L_{\mu\nu\alpha\beta} - \imath N_{\mu\nu\alpha\beta}
\nonumber
\end{eqnarray}
with
\begin{eqnarray}
\nonumber
L_{\mu\nu\alpha\beta} =
\sum_{\gamma} \left[ \right. &x_{\mu\gamma} x_{\gamma\alpha}\delta_{\nu\beta} K^{+}_{\gamma\alpha}&-
\; x_{\mu\alpha} x_{\alpha\gamma} K^{-}_{\beta\nu}- \\
\nonumber
 - & x_{\mu\alpha} x_{\beta\nu} K^{+}_{\mu\alpha}&+
\; x_{\beta\gamma} x_{\gamma\nu} \delta_{\alpha\mu} K^{-}_{\beta\gamma}\left.\right]\\
\nonumber
N_{\mu\nu\alpha\beta} =
\sum_{\gamma} \left[ \right. &x_{\mu\gamma} x_{\gamma\alpha}\delta_{\nu\beta} S_{\gamma\alpha}&-
\; x_{\mu\alpha} x_{\alpha\gamma} S_{\beta\nu}-\\
\nonumber
 -& x_{\mu\alpha} x_{\beta\nu} S_{\mu\alpha}&+
\; x_{\beta\gamma} x_{\gamma\nu} \delta_{\alpha\mu} S_{\beta\gamma}\left.\right]
\end{eqnarray}
\begin{eqnarray}
\nonumber K^{\pm}_{\alpha\beta} &=& K^\pm(\Delta_{\alpha\beta})\,, \;\;\;\;\;\;
S_{\alpha\beta} = S(\Delta_{\alpha\beta})\\
\nonumber K^{\pm}(\Delta) &=& \frac{\pi}{2} I(\Delta) \coth(\frac{\beta \Delta}{2}) \pm \Delta 
\, Y(\Delta)\\
\nonumber S(\Delta)&=& \Lambda \, Y(\Delta)\\
\nonumber Y(\Delta)&=& \gamma_0  \frac{\Lambda^2}{\Lambda^2+\Delta^2}.
\end{eqnarray}

%\vspace{.3cm}

\acknowledgments

We would like to thank E. Calzetta and J.P. Paz for
comments and useful discussions. The work of N. D. A. was
supported by PPARC; F. C. L. and P.I.V. were supported by CONICET, UBA, ANPCyT, and 
Fundaci\'on Antorchas. D.M. is supported by Fundaci\'on Antorchas and CIC.

\end{document}